\documentclass[preprint,10pt]{aastex}
\usepackage{graphicx}
\usepackage{natbib}
\usepackage{graphicx,epsfig}
\usepackage{bm}
\usepackage{color}
\usepackage{comment} 
\usepackage{ulem}

\newcommand{\torresRp}{$5.149\pm0.091$}
\newcommand{\torresRs}{$4.21\pm0.10$} 
\newcommand{\hensRp}{$5.23\pm0.06$} 
\newcommand{\hensRs}{$4.32\pm0.07$}

\begin{document} 

\title{Reanalysis of the Radii 
of the Benchmark Eclipsing Binary V578~Mon}

\author{E.V.\ Garcia\altaffilmark{1,2}, Keivan G.\ Stassun\altaffilmark{1}, Guillermo Torres \altaffilmark{3} }

\altaffiltext{1}{Department of Physics \& Astronomy, 
Vanderbilt University, VU Station B 1807, Nashville, TN 37235, USA} 
\altaffiltext{2}{Fisk-Vanderbilt Masters-to-PhD Bridge Program Graduate Fellow; eugenio.v.garcia@gmail.com}
\altaffiltext{3}{Harvard-Smithsonian Center for Astrophysics; 60 Garden Street, Mail Stop 20, Cambridge,
MA 02138, USA}

\begin{abstract}
V578 Mon is an eclipsing binary system in which both 
stars have masses above 10 M$_\odot$ determined with an accuracy better
than 3\%. It is one of only five such massive eclipsing binaries known 
that also possess eccentric orbits and measured apsidal motions, thus
making it an important benchmark for theoretical stellar evolution models.
However, recently reported determinations of the radii of V578 Mon
differ significantly from previously reported values.
We reanalyze the published data for V578 Mon and
trace the
discrepancy to the use of an incorrect formulation for the 
stellar potentials in the most recent analysis.
Here we report corrected radii for this important benchmark eclipsing binary.
\end{abstract}

\section{Introduction}

In a recent major review of eclipsing binary (EB) stars, \citet{Torres2010} compiled a sample of 95 ``benchmark grade" EBs 
having masses and radii measured with an accuracy of better than $\pm3$\%. This sample of EBs is of fundamental importance to
a broad range of applications in stellar astrophysics. For example, using this sample, \citet{Torres2010} derive empirical 
relations that permit the 
determination of fundamental properties, such as mass and radius, for any star from its directly observable properties, such as effective 
temperature, surface gravity and metallicity. This sample of EBs is moreover important for testing the basic predictions of theoretical stellar 
evolution models, which are in turn used across many areas of research, including star formation, stellar evolution, and exoplanets
(because the inferred properties of the planets generally depend on the assumed properties of the host stars).

One of the 95 EBs included in the \citet{Torres2010} compilation is V578~Mon.
This EB is remarkable by virtue of being one of only nine EBs in the compilation where both stars have masses greater than 10 M$_\odot$,
and of these, is one of only five to have an eccentric orbit and therefore a measurable apsidal motion
\citep{H2000, PavlovskiHensberge2005}. Indeed, the apsidal motion of V578~Mon was only recently measured \citep{Garcia2011}.


We have identified an error in the calculation of the stellar radii
reported in the \citet{Torres2010} compilation. In this brief report, we
trace the source of this error to the use of a formulation for the stellar 
potentials that does not 
correctly include the system's orbital eccentricity. We then provide
updated radii calculated from the correct formulation for the potentials.

\section{Calculation of Stellar Radii for V578 Mon}

 
\cite{Torres2010} reported the 
radii of V578 Mon to be 
$R_{p}=$\torresRp~and $R_{s}=$\torresRs, 
where $R_p$ is the radius of the more massive primary star and
$R_s$ is the radius of the less massive secondary star.
These radii differ from the previously reported values of 
$R_{p} =$\hensRp~and $R_{s}=$\hensRs\ by \citet{H2000}.
While these differences are not highly statistically significant
given the uncertainties quoted in the two studies, the difference is 
slighly more than 1$\sigma$ and any difference is unexpected since 
\citet{Torres2010} recalculated the radii from the same data originally 
presented by \citet{H2000}.

As is standard for the determination of EB radii, the radii are
calculated from the {\it volumetric radius} which is defined as 
$R=(\frac{3}{4\pi}V)^{1/3}$, where $V$ is the volume of the star.
Therefore we suspected the difference may lie in different 
approaches to calculating the stellar volumes, which are themselves
derived from the stellar {\it potential surfaces}. Since \citet{Torres2010}
adopted the same stellar potentials originally reported by \citet{H2000},
the different radii reported by the two authors is likely to be in the
formulation used to transform the potentials to volumes.

\cite{Torres2010} computes $V$ using 
the stellar potentials, $\Omega$, and the stellar mass
ratio, $q \equiv M_2 / M_1$, from \cite{H2000} assuming a circular synchronous orbit. 
More formally,
$V = V(\Omega,q)$ if 
the orbit is circular ($e =0$), 
and if the stars rotate synchronously with the orbit
($F_{1}=F_{2}=1.0$,where $F$ is the ratio of the stellar angular
velocity to the orbital angular velocity). Following
\citet{Kopal89}:  
\begin{equation}
\label{eqn:Omegacirc}
\Omega = \frac{1}{r} + q\{\frac{1}{\sqrt{1-2\lambda r + r^{2}}}-r\lambda \}+\frac{q+1}{2}r^{2}(1-\nu^{2})
\end{equation}
where $\nu = \cos{\theta}$, $\lambda=\cos{\phi}\sin{\theta}$, 
$\theta$ and $\phi$ are instantaneous direction angles between the
two stars, and $r$ is
the distance to the surface of the star from the star's center
in the $(\theta, \phi)$ direction and in units of the instantaneous
separation between the two stars. Through several
complex mathematical operations, equation \ref{eqn:Omegacirc} can be used
to uniquely define $V = V(\Omega,q)$ and therefore $R = R(\Omega,q)$ 
\citep[see Equation 2.18 in][]{Kopal89}.

However, V578~Mon has an eccentric (i.e., non-circular) orbit
with pseudo-synchronous rotation, where $e = 0.0867$, $F_{1}=1.13$ and $F_{2}=1.11$ \citep{H2000}.
Therefore the $R$ cannot be uniquely determined as a function of $\Omega$ and $q$ alone. 
Instead of using equation \ref{eqn:Omegacirc}, 
it is necessary to use $\Omega$ defined for non-circular orbits 
\citep{wilson1979}: 
\begin{equation}
\label{eqn:Omeganocirc} 
\Omega = \frac{1}{r} + q\{\frac{1}{\sqrt{\delta^{2}-2r\lambda\delta+r^{2}}}-\frac{r\lambda}{\delta^{2}}\}+F^{2}\frac{q+1}{2}r^{2}(1-\nu^{2}) 
\end{equation} 
where 
$\delta = D/a$ is the instantaneous separation between the two stars
normalized to the semi-major axis. Equation \ref{eqn:Omeganocirc} 
thus defines $V = V(\Omega,q,F,\delta)$ and 
therefore $R=R(\Omega, q, F,\delta)$. 
Whereas \citet{Torres2010} calculated the V578 Mon radii via
equation \ref{eqn:Omegacirc},
\citet{H2000} uses the Wilson-Devinney code
\citep{wilson1971,wilson1979} which correctly incorporates equation
\ref{eqn:Omeganocirc} in computing the radii. 

As an independent check on the Wilson-Devinney 
 based calculation of the radii, 
we compute $R_{p}$ and $R_{s}$ by solving equation \ref{eqn:Omeganocirc} numerically for $r(\theta,\phi)$ in unit steps of $\theta$ and $\phi$. We set $\Omega_{1}$, $\Omega_{2}$, $F_{1}$, $F_{2}$, $q$ and $e$ to values from \citet{H2000}. 
For noncircular orbits, $\Omega = \Omega(t)$ since the instantaneous separation between the two stars varies over the orbit. We compute the volumetric radius at periastron, setting the instantaneous separation $\delta = 1-e$, since 
the $\Omega$ reported by \citet{H2000}  are the potentials at periastron \citep{wilson1979}. 
We compute $V$ from $r(\theta,\phi)$ discretely as:
\begin{equation} 
\label{eqn:V}
V = \frac{1}{3}\sum\sum r(\theta,\phi)^{3} \sin(\theta) \Delta\phi \Delta\theta
\end{equation}
where $\Delta\phi$ and $\Delta\theta$ are discrete steps in $\phi$ and $\theta$. 

From equation \ref{eqn:V} we compute $R_{p} = 5.24$ and $R_{s}
= 4.33$, which are essentially identical to the radii from \citet{H2000}. 
Finally, in the same fashion we also solve 
for $r(\theta,\phi)$ for the case of a circular orbit
(equation \ref{eqn:Omegacirc}),
and compute $V$ using equation \ref{eqn:V}. 
In this case, we find $R_{p}=5.15$ and $R_{s}=4.21$, 
which are essentially identical
to the radii
reported in \cite{Torres2010}.

Thus, it is clear that the differences in the radii reported for
V578 Mon between \citet{H2000} and \citet{Torres2010} arises from an 
incorrect use by \citet{Torres2010} of the formulation for the stellar
volumes assuming circular, synchronous orbits.

\section{Summary}

We have shown that the radii reported for V578 Mon by 
\citet{Torres2010} are incorrect, with an error of $\sim$2\%
relative to the expected values \citep{H2000}.
While small, the error is unexpected since \citet{Torres2010} used the 
same data as \citet{H2000}, and moreover the highest quality radius 
determinations of EBs should achieve $\sim$1\% precision.
We have traced the error to the use by those authors of the formulation
for the stellar potential surfaces that is 
correct only in the case of circular, synchronous orbits. However,
V578 Mon possesses an eccentric, pseudo-synchronous orbit, and thus
its volumetric radii must be calculated from the formulation for the
stellar potentials that correctly accounts for this \citep{wilson1979}.

Radii for V578 Mon were originally reported by \citet{H2000}, 
who use the Wilson-Devinney code which 
incorporates the correct formulation for the stellar potentials
in an eccentric orbit
(equation \ref{eqn:Omeganocirc}). 
We have performed an independent check 
on the \citet{H2000} radii by solving
equation \ref{eqn:Omeganocirc} numerically; we recover radii 
identical to \cite{H2000}. We are able to recover 
the \citet{Torres2010} radii 
when we solve the equations assuming a circular, synchronous orbit. 

We have checked that the use of the
erroneous formulation for the potential in \citet{Torres2010} does not
significantly alter the radii for any other systems reported in that paper;
fortunately only V578 Mon is sufficiently massive and eccentric to have
been affected.
A forthcoming paper (Garcia et al., in prep.) will report
improved radii of V578~Mon taking into account an updated eccentricity of
$e = 0.07755$ and apsidal motion $\dot\omega = 0.07089$~deg~cycle$^{-1}$
from \citet{Garcia2011}. 
In the meantime we recommend the use of the radii
from \citet{H2000} which we have corroborated here.

\begin{deluxetable}{lrrrc}
\tablecolumns{3}
\tablewidth{0pc} 
\tablecaption{\label{Table:Params} 
V578 Mon Radii Comparisons
}
\tablehead{} 
\startdata
\colhead{} & \colhead{$R_{p}$} & \colhead{$R_{s}$} \\ \hline
 \cite{H2000} & \hensRp  & \hensRs \\ \hline 
\cite{Torres2010} & \torresRp & \torresRs  \\ \hline
This Study & 5.24 & 4.33 \\ \hline
 \\ \hline
\enddata
\end{deluxetable}

\end{document}